# Assessing Consistency of Consumer Confidence Data using Dynamic Latent Class Analysis

SUNIL KUMAR†, ZAKIR HUSAIN ‡ and DIGANTA MUKHERJEE #

*†: Sampling and Official Statistics Unit, Indian Statistical Institute, Kolkata and Alliance University, Bangalore, India (e-mail: sunilbhougal06@gmail.com)*

*‡: Sampling and Official Statistics Unit, Indian Statistical Institute, Kolkata, India (e-mail: digantam@hotmail.com)*

*#: Indian Institute of Technology, Kharagpur, India (e-mail: dzhusain@gmail.com)*

## Abstract

In many countries information on expectations collected through consumer confidence surveys are used in macroeconomic policy formulation. Unfortunately, before doing so, the consistency of responses is often not taken into account, leading to biases creeping in and affecting the reliability of the indices hence created. This paper describes how latent class analysis may be used to check the consistency of responses and ensure a parsimonious questionnaire. In particular, we examine how temporal changes may be incorporated into the model. Our methodology is illustrated using three rounds of Consumer Confidence Survey (CCS) conducted by Reserve Bank of India (RBI).

## I. Introduction

The effect of attitudes of non-financial organizations and consumers on economic activity is a subject of great interest to both policymakers and economic forecasters (see Galstyan and Movsisyan, 2009). In particular, expectations about macroeconomic variables may influence the effectiveness of monetary and fiscal policy and also the direction of the economy (see Phelps, 1968). This underlines the importance of incorporating such expectations in the process of formulation of macroeconomic policy. Information about such expectations is collected by policy makers in many countries through consumer confidence surveys. Examples of such surveys are Business Tendency and Consumer Surveys of the Organization for Economic Co-operation and Development in Armenia, The Nielsen Global Survey of Consumer Confidence and Spending Intentions in Hong Kong, Westpac-Melbourne Institute Survey of Consumer Sentiment, etc. Based on such surveys, consumer confidence indices are constructed and often used in policy making as a proxy for consumers' expectations about the future trend of the economy.

In India, information about the households' perceptions about the current economic situation and their expectations about future economic changes is collected on a quarterly basis by Reserve Bank of India since June 2010 through the CCSs. The responses to these surveys are analyzed to obtain pictures of households' opinions on the overall economic situation (current and future) and their material security (current and future). In addition, the RBI calculates indices of current and future conditions (see RBI Bulletin, 2012). The survey data are a potentially useful tool to monitor temporal changes in households' expectations. Accordingly, the survey results are used by the RBI to formulate monetary policy and determine key monetary variables like Cash Reserve Ratio, interest rates, etc.

Unfortunately, while constructing such indices, the reliability and consistency of responses is often not taken into account (see Katona, 1946, 1947). This leads to overlooking of biases in consumer responses, and may even affect the reliability of the indices. The RBI, too, uses the survey response to form monetary policy variables without examining the consistency of responses. This paper describes how latent class analysis (LCA)—a latent variable model with discrete latent and indicator variables—may be used to check the consistency of responses and identify the variables that may be used to construct a reliable index of consumer sentiment about the economy. Our methodology is illustrated using the CCS conducted by RBI.

The paper is structured as follows. Section 2 describes the framework of the LCA model. Sections 3 and 4 present the analysis of three rounds of CCS data using the software application poLCA.[1] Finally, in section 5 and 6, we summarize our findings and identify potential areas for further research.

## II.    LCA models and response biases

### LCA Models

LCA is a statistical method for clustering the related cases (identifying latent classes) from multivariate categorical data, pioneered by Lazarsfeld (1950) and Lazarsfeld and Henry (1968). LCA models do not rely on traditional modeling assumptions like normal distribution, linear relationship, homogeneity etc. Both the latent and indicator variables of this model are discrete. LCA is a subset of structural equation modelling, used to classify unobservable sub-groups or subtypes of cases in multivariate categorical data. As in factor analysis, the LCA can also be used to group cases according to their maximum likelihood class membership. LCA may be applied to classify types of attitude structures from survey responses, consumer segments from demographic and preference variables, or categorize subpopulations based on their responses to test items. LCA has the advantage of making no assumptions about the distribution of the indicators other than that of local independence which says that the indicators share a common latent variable, but the errors in measurement are uncorrelated. Subsequent development allows this assumption to be relaxed. Even when we have dependence between indicators, it can be incorporated via interaction term among the indicator variables (e.g. Harper, 1972; Vacek, 1985; Hangenaars, 1988; Espeland and Handelman, 1989; Sinclair and Gastwirth, 1996; Reboussin *et al.*, 2008; Bertrand and Haftner, 2011).

---

[1] The analysis is based on the most complete and most user-friendly package for the estimation of latent class models and latent class regression models in R (see Linger and Lewis, 2011; R Development Core Team, 2010)

**Applications of LCA**

Clogg and Goodman (1984) were the first to introduce a latent class model in which class (unobservable categories of the latent variable) membership probabilities and item response probabilities are conditioned on membership in an observed group. LCA with covariates extends the basic LCA model to include predictors of class membership. In this extension, latent class membership probabilities are predicted by covariates through a logistic link (e.g. Bandeen et al 1997; Dayton and Macready 1988). Latent class models have been applied in many domains. For example, Sullivan and Kessler (1998) used LCA to determine empirically the typologies of depressive symptoms in the national co-morbidity survey (NCS) in U.S. Yan *et al.* (2012) applied latent class models to the data from two cycles of the National Survey of Family Growth (NSFG).

Another application of LCA models is to test for the presence of different type of response biases in consumer surveys. For instance, surveys may contain biases like acquiescence, extreme response style and social desirability:

(i)   Acquiescence describes the general tendency of a person to provide confirmatory answers to items of questionnaires heedlessly of the content of the items (see Messick, 1967).

(ii)  Extreme response bias refers to the tendency to uniformly endorse an optimistic view about the economy (positive bias) or uniformly report a negative perception about the economy (negative bias), irrespective of the information content sought (see Baumgartner and Steenkamp, 2006).

(iii) Social desirability is often referred as tendency to responding in a way that presents them favorably according to current cultural norms (see Mick, 1996).

Identification of the sources of such errors is a prerequisite to using results of the consumer confidence surveys in formulating macroeconomic policies. The most promising way of accounting for such biases is through application of statistical techniques to analyze survey responses. Given the qualitative nature of the data it is not possible to use the classical least squares methodology. So the more flexible alternative, LCA, has been used in such exercises. For instance, Biemer and Winsen (2002) used LCA for estimating classification errors. This led to identification of errors in the design of questionnaire and wording of questions (National Household Survey of Drug Abuse), necessitating adjusting estimates of prevalence of drug use for classification error bias. Bialowolski (2012) used LCA to examine the sentiment bias and extreme response bias on the individual data from the State of Households' inflation expectation survey conducted in Poland.

The present paper is part of an ongoing exercise to use LCA models to analyze the pattern of misclassification arising out of response biases in Reserve Bank of India's Consumer Confidence Survey. Specifically, we propose to test for the presence of extreme response bias in the survey and identify the items affected by this bias. It extends an earlier analysis undertaken for a single time period to show how the impact of temporal changes may be incorporated into the LCA model to evaluate reliability of CCS responses and identify sources of biases. Since the survey is conducted on a quarterly basis we have taken the data for the same quarter (first quarter in the financial year, ending on $30^{th}$ June) in three successive years —2010, 2011 and 2012. These were the $1^{st}$, $5^{th}$ and $9^{th}$ round of the series resp. This helps to factor out any seasonal effect between rounds.

## III. Database and methodology

### Database

The Consumer Confidence Survey (CCS) is a household survey designed to measure an assessment of the consumer sentiments of the respondents based on their perceptions of the general conditions and their own financial situation. The CCS design is a multistage stratified sample covering six metropolitan cities, viz., Bengaluru, Chennai, Hyderabad, Kolkata, Mumbai and New Delhi. Each city is stratified into three major areas and each major area is further stratified into three sub-areas. From each sub-area, about 100 respondents are selected randomly. For each round of survey, 5400 respondents are selected (900 respondents from each city).The assessments are made in two parts, viz., current situation as compared with a year ago and expectations for a year ahead. The survey schedule consists of qualitative questions pertaining to impression about economic conditions, views on household circumstances, perceptions on price level and employment prospects and developments in real estate prices and views on growth potential of the Indian economy (see Table1). In addition, information was elicited on age,

gender, occupation and annual income of respondent. For this study, we have taken the data for

the same quarter (first quarter in the financial year, corresponding to June) in three successive

years —2010, 2011 and 2012, in the analysis. Data from a total of 15523 schedules were used

(about 4.2 percent of total responses were found to be unsuitable by RBI and not released).

TABLE 1

*Set of questions from the standardized Consumer Confidence Survey Questionnaire which is used in our analysis*

| Question number and Code | Question wording | | Answer Categories (representing also scale points) |
|---|---|---|---|
| Q1 (A) | How do you think economic conditions have changed compared with one year ago? | 1<br>2<br>3 | Have improved<br>Have remained the same<br>Have worsened |
| Q12 (B) | How do you think the overall prices of goods and services have changed compared with one year ago? | 1<br>2<br>3 | Have gone up<br>Have remained almost unchanged<br>Have gone down |
| Q5 (C) | What do you think about your household circumstances compared with one year ago? | 1<br>2<br>3 | Have become somewhat better off<br>Difficult to say<br>Have become somewhat worse off |
| Q6 (D) | How has your income (or other family members' income) changed from one year ago? | 1<br>2<br>3 | Has increased<br>Has remained the same<br>Has decreased |
| Q8 (E) | How have you (or other family members') changed consumption spending compared with one year ago? | 1<br>2<br><br>3 | Have increased<br>Have neither increased nor decreased<br>Have decreased |
| Q3 (F) | How do you foresee economic conditions one year from now? | 1<br>2<br>3 | Will improve<br>Will remain the same<br>Will worsen |
| Q13 (G) | In which direction do you think prices will move one year from now? | 1<br>2<br>3 | Will go up<br>Will remain almost unchanged<br>Will go down |
| Q7 (H) | What do you expect your income (or other family members' income) will be one year from now? | 1<br>2<br>3 | Will increase<br>Will remain the same<br> Will decrease |
| Q10 (I) | Do you plan to increase or decrease your spending within the next twelve months? | 1<br>2<br>3 | Increase<br>Neither increase nor decrease<br>Decrease |
| Q11 (J) | In consideration of the situation over the next twelve months, are you worried about your (or other | 1<br>2<br>3 | Not particularly worried<br>Slightly worried<br>Quite worried |

family members') employment or working arrangements (pay. Job position, and benefits) at the current workplace?

---

*Source:* Consumer Confidence Survey, conducted by Reserve Bank of India.

## Research methodology

The basic structure of LCA was developed by Lazarsfeld (1950), Lazarsfeld and Henry (1968), Goodman (1974), Clogg (1995), Clogg and Manning (1996) among others. A characteristic feature of LCA is that the latent variables and items (directly observed manifest variables) are discrete. The relation between the latent variable and its indicators is not deterministic, but probabilistic. LCA produces the classification probabilities of the individual respondents, conditional on the indicator variables. For example $\pi_{a|x}^{A|X}\pi_{a|x}^{A|X}$ is the conditional probability $P(A=a|X=x)$ where X is the latent variable and A is one of the indicator variable.

The type of models considered in our analysis is limited to simple extensions of the basic latent class model as proposed in Biemer and Winsen (2002) with grouping variables defined by annual income (Z) and time (T). The use of grouping variables has been suggested by Hui and Walter (1980) to either ensure that the model is identifiable, or to improve the fit of the model or to reduce the effect of unobserved heterogeneity. Although our model is identifiable, the remaining two issues remain a matter of concern. Following the Hui-Walter approach, we choose income as the grouping variable. This is justified on two grounds.

    a) An essential assumption of the LCA model is that the error probabilities are the same for each individual in the population. When this assumption is violated, as is likely

for groups formed on the basis of economic variables like income, we say that the population is heterogeneous with respect to the error probabilities. Suppose however that there is a categorical variable (H) such that, within the categories of H, the assumption of homogeneity holds. When H is known then adding the term XHA to the model will achieve conditional homogeneity for XA, which is sufficient for model validity (Biemer, 2011).

b) Unlike other possible grouping variables like age, gender, occupation and household size, income is highly correlated with the latent variable (consumer confidence about the economy).

The following is a description of the models that were used in the analysis

(i)   Model 0: Assumes latent class membership affected by annual income (Z).

(ii)  Model 1: Assumes latent class membership affected by occupation (O).

(iii) Model 2: Assumes latent class membership affected by both annual income (Z) and occupation (O).

(iv)  Model 3: Assumes latent class membership affected by annual income (Z) and economic condition (A and F i.e. retrospective and prospective).

All the above defined models postulate that the consumer confidence varies across different combinations of (Z, T) values. Model 0 reflects a type of dependence of the error terms on the grouping variables but not the full dependence represented in other models. Models 1, 2 and 3 represent the three possible cases of full dependence between the indicator variables, has been discussed in detail by Hagenaars (1988). Given the relatively large number of observed variables measuring the latent variable and the number of response categories per variable, the number of parameters is fairly high. For this reason larger models were not considered.

After identifying the appropriate model from the above, we undertake the following analysis:

(i)    Assess the optimal number of response classes for the latent variable;

(ii)    Identify indicators for which responses are inconsistent. We first ran the program for the three rounds separately, with all indicators A through J. We found that the performance of indicator variables B and G is uniformly poor in the sense of yielding a high proportion of inconsistent responses in all the three rounds of data. Therefore, to reduce the extreme probabilities of indicator variables in the pooled analysis (with all three rounds of data together), we ran the program without the indicator variables B and G;

(iii)    Verify consistency of responses for remaining part of the RBI questionnaire, after dropping the inconsistent indicator variables; and,

(iv)    Estimate the contribution of each of the remaining indicator variables to the latent or outcome variable (viz. level of consumer confidence about the economy).

This will ensure a questionnaire that is parsimonious in terms of number of items and elicits responses which are consistent.

## IV.   Analysis

**Performance of Alternative models**

The first step of our analysis is selection of the 'best' model from the list of four models formulated. Table 2 reports necessary statistics for comparing between the four alternative models. Selection is on the basis of Bayesian Information Criterion (BIC) values as suggested by Lin and Dayton (1997). From Table 2 we can see that Akaike Information Criterion (AIC) and BIC are minimized in Model 3. Further, the Log Likelihood (LL) value is also satisfactory for this model. Therefore Model 3 is used in our subsequent analysis.

TABLE 2

*Model diagnostics for alternative classification error models*

| Model | Degrees of freedom | Number of parameters | Log Likelihood (LL) Value | AIC | BIC |
|-------|-------------------|---------------------|--------------------------|-----|-----|
| Model 0 | 6504 | 56 | -96096.55 | 192305.1 | 192733.5 |
| Model 1 | 6496 | 64 | -93670.47 | 187468.9 | 187958.5 |
| Model 2 | 6496 | 64 | -92482.53 | 185093.1 | 185582.7 |
| Model 3 | 6488 | 72 | -91253.31 | 182650.6 | 183201.4 |

**Considering alternative number of classes**

A latent class is a variable indicating underlying subgroups of individuals based on observed characteristics. Membership in the subgroup is said to be "latent" because membership in a class cannot be directly observed. In the context of CCS data for instance, classes indicate the number of categories into which the responses about perception about current scenario of the Indian economy and future anticipated trend of the economy may be divided. In this study we consider three alternatives:

(i) Two-class (where consumers rate current scenario/future prospects of the economy as either positive or negative)

(ii) Three-class (where consumers rate current scenario/future prospects of the economy as either positive, indifferent or negative)

(iii) Four-class (which incorporates strength of confidence about current situation/future scenario of the economy).

We next try to identify which class is appropriate under Model 3 by undertaking LCA on the three alternate versions of Model 3. The full results of the LCA for 2- and 4-classes of model 3 are omitted, presenting only the goodness of fit statistics for 2-, 3- and 4-classes of the model 3 in Table 3. Both the AIC and BIC criteria indicate that the three class model is the most suitable for analysis in terms of goodness of fit and parsimony.

TABLE 3

*Goodness of fit results for 2-, 3- and 4-classes of Model 3*

| Number of Classes (n) | n = 2 | n = 3 | n = 4 |
|---|---|---|---|
| Estimated n-class population shares | 0.5293  0.4707 | 0.4043  0.3507  0.2450 | 0.5585  0.3175  0.1100  0.0141 |
| Predicted n-class memberships (by modal posterior prob.) | 0.5333  0.4667 | 0.4084  0.3486  0.2450 | 0.5585  0.3175  0.1100  0.0141 |
| Number of observations | 15523 | 15523 | 15523 |
| Number of estimated parameters | 44 | 72 | 100 |
| Residual degrees of freedom | 6516 | 6488 | 6460 |
| Maximum log-likelihood | -97542.45 | -91253.31 | -106488.7 |
| AIC | 195172.9 | 182650.6 | 213177.4 |
| BIC | 195509.5 | 183201.4 | 213942.4 |
| $\chi^2$ (Chi-square goodness of fit) | 172918.5 | 65663.3 | 582888.7 |

**Interpretation of three-class Model 3**

Table 4 illustrates the estimated class conditional response probabilities for the indicators A, C, D, E, F, H, I and J, with each row corresponding to a latent class of consumer sentiment, and each column corresponding to classes of the indicator variable values. Thus, the first row (class 1) reports the analysis results for respondents who are in general optimistic about the economy, the second row (class 2) reports results for respondents who feel that the economic prospects have remained unchanged, while the third row states results for respondents with negative perceptions in general about the economy — as classified by the indicator variable. For each row, the columns give conditional probabilities that the respondent has a positive/indifferent/negative feeling (respectively) in terms of the latent variable. For example, second row, first column of each block is the conditional probability P(A= indifferent|X= improve). In particular, the diagonal cells (a=x) give the probability of consistent responses. For indicator A, then values are P(A = positive| X = improve) = 1; P(A = indifferent| X = remain same) = 0.3635 and P(A = negative| X = deteriorate) = 0.6764. Similarly we interpret the results for other indicators.

TABLE 4

*Classification probabilities for the three class model 3*

Conditional item response (column) probabilities, by Indicator variable, for each class (row).

| Indicator variables | Conditional item responses (X) | | |
|---|---|---|---|
| | Pr(1 = improve) | Pr(2 = remain same) | Pr(3 = deteriorate) |
| $A = Economic condition, retrospective | | | |
| Class 1: Positive | 1 | 0 | 0* |
| Class 2: Indifferent | 0.2996 | 0.3635 | 0.337 |
| Class 3: Negative | 0.1766** | 0.1471 | 0.6764 |
| $C = Household circumstances | | | |
| Class 1: Positive | 0.8407 | 0.1016 | 0.0578* |
| Class 2: Indifference | 0.52 | 0.3836 | 0.0964 |
| Class 3: Negative | 0.0107** | 0.0858 | 0.9035 |
| $D = Income, retrospective | | | |
| Class 1: Positive | 0.9284 | 0.0463 | 0.0253* |
| Class 2: Indifference | 0.582 | 0.3997 | 0.0184 |
| Class 3: Negative | 0.0218** | 0.1947 | 0.7835 |
| $E = Spending, retrospective | | | |
| Class 1: Positive | 0.9129 | 0.0691 | 0.018* |
| Class 2: Indifference | 0.6466 | 0.2924 | 0.0609 |
| Class 3: Negative | 0.4332** | 0.1028 | 0.4639 |
| $F = Economic condition, prospective | | | |
| Class 1: Positive | 1 | 0 | 0* |

| Indicator variables | Conditional item responses (X) | | |
|---|---|---|---|
| | Pr(1 = improve) | Pr(2 = remain same) | Pr(3 = deteriorate) |
| | 0.3574 | 0.4495 | 0.1931 |
| | 0.2016** | 0.2913 | 0.5072 |
| $H = Income, prospective | | | |
| Class 1: Positive | 0.907 | 0.0851 | 0.0079* |
| Class 2: Indifference | 0.5556 | 0.4309 | 0.0135 |
| Class 3: Negative | 0.2674** | 0.435 | 0.2976 |
| $I = Spending, prospective | | | |
| Class 1: Positive | 0.8041 | 0.1463 | 0.0496* |
| Class 2: Indifference | 0.5376 | 0.391 | 0.0714 |
| Class 3: Negative | 0.3078** | 0.3579 | 0.3344 |
| $J = Workplace issues | | | |
| Class 1: Positive | 0.6319 | 0.2502 | 0.1179* |
| Class 2: Indifference | 0.4718 | 0.4001 | 0.1281 |
| Class 3: Negative | 0.1517** | 0.3294 | 0.5189 |

* (**) indicate Extreme false negative (positive) probability of all indicator variables.

Inconsistent responses occur when x ≠ a. The probabilities of error in classification for the indicator variable AA is given by $\pi_{a|x}^{A|X} \pi_{a|x}^{A|X}$ for a ≠x. Extreme response biases will occur when we observe one of the following two scenarios:

(i)    Case of extreme false positive probability = P(A= negative| X = improve); indicated by * in table 3.

(ii)     Case of extreme false negative probability = P(A= positive| X = deteriorate); indicated by ** in table 3.

The extreme false positive probability for A is $\pi^{A|X}_{a=1|x=3} = 0.1766$; $\pi^{A|X}_{a=1|x=3} = 0.1766$ and the extreme false negative probability for A is $\pi^{A|X}_{a=3|x=1} = 0.0000$; $\pi^{A|X}_{a=3|x=1} = 0.0000$. Similarly, $\pi^{C|X}_{c=1|x=3} = 0.0107$; $\pi^{C|X}_{c=1|x=3} = 0.0107$ and $\pi^{C|X}_{c=3|x=1} = 0.0578$; $\pi^{C|X}_{c=3|x=1} = 0.0578$ are the respective extreme false positive and extreme false negative probability for CC, and so on.

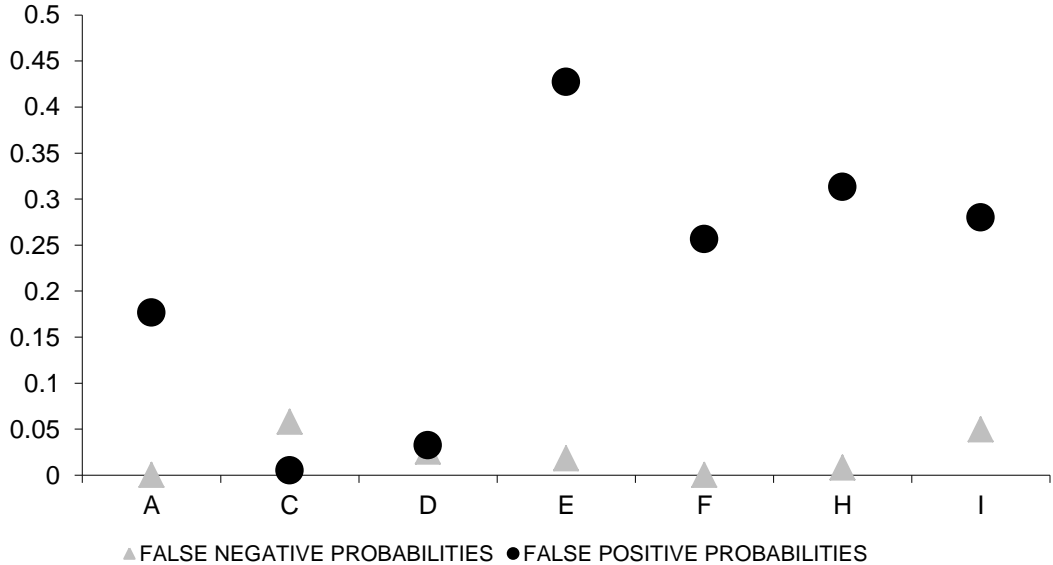

Figure 1.Extreme false negative and extreme false positive probabilities for different indicators

Information about extreme false positive and negative probabilities from Table 4 is summarized in Figure 1. Figure 1 illustrates that the extreme false positive probabilities are all below 0.45 and extreme false negative probabilities are close to zero.

We also calculate the probabilities of consistent classification for all indicators separately with the estimated class population shares. This probability is calculated as follows:

$$P_{JCC} = P(consistent\ classification\ for\ indicator\ vairable\ 'J'in\ all\ classes\ \forall J\ )$$

$$= P \begin{pmatrix} i \in n | i \in group\ '\sigma'by\ class\ population\ shares \\ and\ for\ indicator\ variable\ j\ \forall J;\ \sigma = 1, 2, 3 \end{pmatrix}$$

Similarly, the probability of misclassification for extreme cases is defined as

$$P_{JCC}^* = P \begin{pmatrix} misclassification\ for\ all\ indicators\ with\ estimated\ class\ population \\ shares\ and\ for\ indicator\ variable\ J\ with\ \boldsymbol{extreme}\ cases \end{pmatrix}$$

$$= P \begin{pmatrix} i \in n | i \in group\ '\sigma'by\ class\ population\ shares \\ and\ for\ extreme\ cases\ of\ all\ indicators \end{pmatrix} =$$

$$P \begin{pmatrix} i \in n | i \in group\ '\sigma'by\ class\ population\ shares \\ and\ for\ extreme\ cases\ of\ all\ indicators \end{pmatrix}.$$

A comparison of these two probabilities is important to determine whether the consumers' sentiment is being captured consistently through the indicators. The probabilities are given in Figure 2.

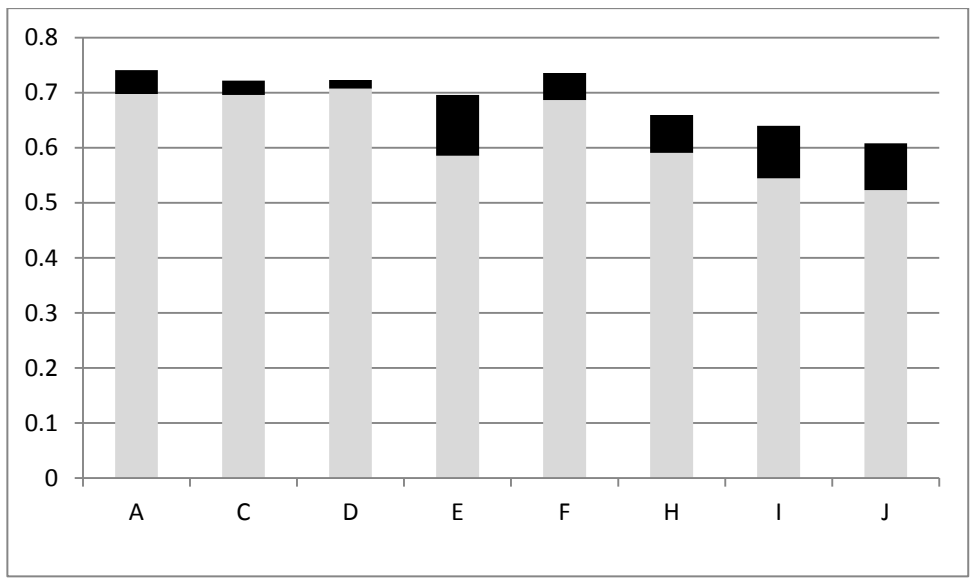

Figure 2.Consistent classification and misclassification probabilities of all indicators

The height of the grey bar gives the probability of consistent classification of each indicator while the height of the black bar gives the probability of inconsistent classification for that indicator. We can see that the probability of consistent classification of all indicators is much higher than the probability of misclassification. From results stated in Table A1 in Appendix II, we find that all the indicators are performing well in terms of classification. The correct classification percentage for A, C, D and F are all close to 70% while that for the other are in the range 50 to 60 %. The misclassification percentage is uniformly low (not more than 11% as for E). On average, the probability of consistent misclassification is 0.062 and consistent classification is 0.629. Therefore, the probability of inconsistent responses for individual indicators is within acceptable limits.

## V.    Contribution of different indicators towards classification

Our analysis in section 4 shows that the responses are consistent, with probabilities of inconsistent responses to individual indicators lying within acceptable limits. Given this consistency, the next step is to estimate the contribution of each indicator to the overall classification indicator, and examine the statistical significance of the contributions.

Table 5 reports the significance of all the indicators included in the interaction terms of the model (A & F) both in terms of their direct effect as well as the interaction effect. Both these indicators as well as the grouping variable Z (income) are significant contributors to the classification exercise. The interaction of the indicators with grouping variable and time (two-way and three-way) are significant. The only exception is the interaction terms of Z and time for class 2 versus class 1.

The effect of time is uniformly significant individually as well as in the interactions, indicating that perceptions are changing over time but doing so in a predictable manner. This justifies the incorporation of the time dimension in all models, which also enhance the predictive power.

TABLE 5

*Coefficients of indicators for 3 latent classes, Model 3*

| 2/1 | Class 2 Vs Class 1 | | | |
|---|---|---|---|---|
| | Coefficient | Std. error | t value | Pr(>|t|) |
| (Intercept) | -16.08575 | 0.42613 | -37.748 | 0.00 |
| A | 9.6962 | 0.2121 | 45.715 | 0.00 |
| F | 6.34009 | 0.21573 | 29.389 | 0.00 |
| Z | -5.87079 | 0.23865 | -24.6 | 0.00 |
| T | -2.97835 | 0.18605 | -16.009 | 0.00 |
| A:Z | 2.56177 | 0.1186 | 21.601 | 0.00 |
| F:Z | 2.49952 | 0.1216 | 20.555 | 0.00 |
| A:T | 0.60433 | 0.08627 | 7.005 | 0.00 |
| F:T | 1.62646 | 0.08682 | 18.734 | 0.00 |
| Z:T | -0.1905 | 0.10532 | -1.809 | 0.07 |
| A:Z:T | -0.30975 | 0.04843 | -6.396 | 0.00 |
| F:Z:T | 0.69873 | 0.04958 | 14.094 | 0.00 |
| 3/1 | Class 3 Vs Class 1 | | | |
| (Intercept) | -19.4534 | 0.49566 | -39.247 | 0.00 |
| A | 10.92881 | 0.1778 | 61.468 | 0.00 |
| F | 6.54522 | 0.18229 | 35.905 | 0.00 |
| Z | -8.64174 | 0.28751 | -30.057 | 0.00 |
| T | -2.05824 | 0.19549 | -10.528 | 0.00 |
| A:Z | 2.65332 | 0.09553 | 27.776 | 0.00 |
| F:Z | 3.4244 | 0.09598 | 35.677 | 0.00 |
| A:T | 0.22031 | 0.08278 | 2.661 | 0.01 |
| F:T | 1.72565 | 0.0807 | 21.383 | 0.00 |
| Z:T | 0.67605 | 0.11395 | 5.933 | 0.00 |
| A:Z:T | -0.32845 | 0.04516 | -7.273 | 0.00 |
| F:Z:T | 0.35442 | 0.04412 | 8.034 | 0.00 |

## VI. Conclusion

In many countries expectations collected through consumer confidence surveys are used in the process of macroeconomic policy formulation. Unfortunately, before doing so, the reliability of responses is often not taken into account, leading to biases creeping in and affecting the reliability of the indices. In particular extreme response bias is an important category of bias that may distort the results of the survey. Using three rounds of Consumer Confidence Survey conducted by Reserve Bank of India this paper describes how latent class analysis may be used to check the reliability of responses and evaluate the performance (in terms of consistency and contribution to the overall outcome indicator, viz. consumer confidence about the economy) of eight indicators used in the survey.

The goodness of fit results (using likelihood ratio, AIC and BIC criteria) all indicate that the optimal model classifies the responses into three classes, reflecting the attitude of responses about the economy (positive, indifference and negative). This is consistent with current RBI practice. Indicators B and G had to be dropped from our analysis because of severe inconsistencies in responses in all three rounds. We suggest that these questions should either be reformulated (and the consistency of responses rechecked), or they should be dropped from the questionnaire altogether. The results show that all the factors used in the analysis are significant at the individual level and also interacting with time and income groups. The classification performance is uniformly good for these indicators. Thus, there is no need to further reduce the questions in the CCS.

The optimal configuration of interactions among the indicators and the grouping factors like time and income is also identified. The attitude of consumers reveals a predictable change over time as well as over income groups. Thus differential effects of alternative policies may be judged from our models. Hence, our results may help in shifting from broad based policies to policies with greater potential for targeting specific groups across socio economic lines. Secondly, such policies may also incorporate the impact of temporal changes in attitude or preferences, making such policies more dynamic. Thus, it would be interesting to study such a scenario with the help of a panel data where we would be able to identify the impact of time on changes in consumer preferences and outlook.

**Appendix II:**

TABLE A1

*Consistent classification and misclassification probabilities for all indicators*

| Indicator Variable | P(Consistent Classification) | P(Misclassification) |
|---|---|---|
| A | 0.697497 | 0.043267 |
| C | 0.695781 | 0.02599 |
| D | 0.707484 | 0.01557 |
| E | 0.585286 | 0.1105 |
| F | 0.686204 | 0.049392 |
| H | 0.590729 | 0.068707 |
| I | 0.544149 | 0.095464 |
| J | 0.522923 | 0.084833 |
| Average | 0.628757 | 0.061715 |